\documentclass[conference]{ieeeconf}
\IEEEoverridecommandlockouts

\usepackage{cite}
\usepackage{amsmath,amssymb,amsfonts}
\usepackage{algorithmic}
\usepackage{graphicx}
\usepackage{textcomp}
\usepackage{xcolor}
\usepackage{cuted}
\usepackage{siunitx}
\usepackage{tikz,pgfplots}
\usetikzlibrary{calc}
\def\BibTeX{{\rm B\kern-.05em{\sc i\kern-.025em b}\kern-.08em
    T\kern-.1667em\lower.7ex\hbox{E}\kern-.125emX}}

\pgfplotsset{compat=1.18} 

\usepackage[font=footnotesize, labelsep=period]{caption}
\begin{document}

\title{\LARGE \bf
Real-Time Non-Smooth MPC for Switching Systems: \\ Application to a Three-Tank Process
}

\author{Hendrik Alsmeier\textsuperscript{1},
Felix Häusser\textsuperscript{1},
Andreas Knödler\textsuperscript{1},
Armin Nurkanović\textsuperscript{2},
Anton Pozharskiy\textsuperscript{2},\\
Moritz Diehl\textsuperscript{2},
and Rolf Findeisen\textsuperscript{1}%
\thanks{\textsuperscript{1}Control and Cyber-Physical Systems Laboratory (CCPS),
Technical University of Darmstadt, Germany. \{hendrik.alsmeier, felix.haeusser, rolf.findeisen\}@iat.tu-darmstadt.de, andreas.knoedler@mail.de}
\thanks{\textsuperscript{2}Department of Microsystems Engineering (IMTEK),
University of Freiburg, Germany. \{armin.nurkanovic, anton.pozharskiy, moritz.diehl\}@imtek.uni-freiburg.de}
\thanks{This research was supported by DFG via projects 504452366 (SPP~2364) and 525018088,
by BMWK via 03EI4057A and 03EN3054B, and by the DLR via 20E2219B.}}

\maketitle


\begin{abstract}
Real-time model predictive control of non-smooth switching systems remains challenging due to discontinuities and the presence of discrete modes, which complicate numerical integration and optimization. 
This paper presents a real-time feasible non-smooth model predictive control scheme for a physical three-tank process, implemented without mixed-integer formulations. 
The approach combines Filippov system modeling with finite elements and switch detection for time discretization, leading to a finite-dimensional optimal control problem formulated as a mathematical program with complementarity constraints. 
The mathematical program is solved via a homotopy of smooth nonlinear programs. 
We introduce modeling adjustments that make the three-tank dynamics numerically tractable, including additional modes to avoid non-Lipschitz points and undefined function values. 
Hardware experiments demonstrate efficient handling of switching events, mode-consistent tracking across reference changes, correct boundary handling, and constraint satisfaction. 
Furthermore, we investigate the impact of model mismatch and show that the tracking performance and computation times remain within real-time limits for the chosen sampling time.
The complete controller is implemented using the non-smooth optimal control framework NOSNOC.
\end{abstract}


\section{Introduction}
\label{sec:introduction}
Many real-world systems combine continuous physical dynamics with discrete switching events, for example, due to valve operations in process units or contact and friction phenomena in mechanical systems. Such behavior is naturally described by models that are only piecewise smooth, exhibiting non-smooth dynamics at region boundaries~\cite{acary2008}.

Model predictive control (MPC) is a well-established and reliable control strategy that profits from the use of high-fidelity models~\cite{lucia2016,rawlings2017}. However, for systems that are only piecewise smooth, the discontinuities at region boundaries pose challenges for numerical integration and optimization, significantly increasing computational complexity~\cite{acary2008}.


Commonly, switching behavior in MPC is handled by encoding the modes explicitly using the mixed logical dynamical (MLD) framework~\cite{bemporad1999,bemporad2002}, which leads to mixed-integer optimal control problems. This formulation preserves the switching structure by design and has been extensively studied in terms of modeling, stability, and applications. However, solving mixed-integer problems online is computationally demanding for fast sampling times and nontrivial horizons~\cite{bemporad1999,bemporad2002, Villa2003,lazar2006}, and nonlinearities further increase the number of integer variables required. Recent progress has improved warm-started branch-and-bound methods for hybrid MPC, including results on a three-tank benchmark~\cite{yaakoubi2023}.

In addition to mixed-integer formulations, continuous-time approaches based on non-smooth analysis and Filippov theory provide a rigorous description of switching dynamics. However, standard smooth-MPC formulations are not well suited for such systems: sensitivities become ill-defined near switching boundaries, and numerical integration may fail to preserve the underlying mode changes~\cite{liberzon2003switching,rawlings2017}.
In practice, preserving the switching logic while achieving real-time feasibility with acceptable model accuracy is often challenging. Smoothing or regularization can improve tractability but generally degrades constraint satisfaction and yields inconsistent sensitivities at switching boundaries~\cite{Nurkanovic2020,Stewart2010}. 
The non-smooth optimal control framework (NOSNOC) introduced in~\cite{Nurkanovic2022} uses finite elements with switch detection (FESD)~\cite{Nurkanovic2024} to discretize non-smooth systems. This transcribes the resulting optimal control problems into mathematical programs with complementarity constraints (MPCCs). 
The arising MPCCs can be solved efficiently using regularization–homotopy methods, cf.~\cite{Nurkanovic2024b}. 
The FESD approach resolves sensitivity issues near switching boundaries by aligning discretization nodes with switching events, keeping the active set constant within each finite element, and solving the resulting complementarity-constrained nonlinear programs (NLPs) without using integer variables~\cite{Nurkanovic2024,Nurkanovic2024a}.

\textit{Contributions.} 
In this work, we present an MPC formulation based on the NOSNOC framework and report, to the best of our knowledge, the first real-time closed-loop implementation on a physical three-tank system exhibiting switching behavior due to inter-tank flows and drains. 
The proposed controller preserves the switching logic and implicitly detects mode transitions via the FESD discretization, ensuring consistent mode sequencing and correct sliding behavior without relying on mixed-integer formulations. 
We demonstrate closed-loop tracking performance, constraint satisfaction, and real-time computation under varying references, confirming the feasibility of a continuous-optimization pipeline from CasADi~\cite{casadi} through NOSNOC to IPOPT~\cite{wachter2006}. 

\textit{Outline.}
Section~\ref{sec:problem} formalizes the piecewise-smooth model and Filippov inclusion for the three-tank system. 
Thereafter, Section~\ref{sec:nosnoc_mpc} presents the NOSNOC MPC formulation, the FESD discretization, and the resulting MPCC. 
Section~\ref{sec:results} reports hardware experiments and simulation studies, and Section~\ref{sec:conclusion} concludes with limitations and directions for further development.

\section{Problem Setup}
\label{sec:problem} 


\noindent We consider a continuous-time nonlinear system
\begin{equation}
\dot{x}(t)=f(x(t),u(t)),\label{eq:general_system}
\end{equation}
with $x(t)\in\mathbb{R}^{n_x}$, $u(t)\in\mathbb{R}^{n_u}$. For brevity, we will forgo the time dependents for $x$ and $u$ in the following if it is not specifically relevant.
In this work, $f:\mathbb{R}^{n_x}\times\mathbb{R}^{n_u}\to\mathbb{R}^{n_x}$ is piecewise-smooth function. Thus, we assume there exist disjoint, nonempty, connected, locally finite, open regions
$\{R_i\}_{i=1}^{m}$ with closure $\overline{R}_i$ and piecewise-smooth boundaries $\partial R_i$  such that
$\bigcup_{i=1}^{m} R_i = \mathbb{R}^{n_x} \setminus \Gamma$ with $\Gamma := \bigcup_{i=1}^m \partial R_i $ of Lebesgue measure 0. For each mode $i\in\{1,\ldots,m\}$, let
$f_i:\mathbb{R}^{n_x}\times\mathbb{R}^{n_u}\to\mathbb{R}^{n_x}$ be smooth on an open neighborhood of $\overline{R}_i$. Thus, we define the piecewise-smooth system $f$ as~\cite{Nurkanovic2022}
\begin{equation}
    f(x,u) := f_i(x,u)\quad \text{if } x\in R_i,\; i\in\{1,\ldots,m\}.
    \label{eq:PSS}
\end{equation}
In general, the dynamics are not defined on the boundaries $\partial R_i$, which may introduce discontinuities. 
This presents issues for model-based and gradient-based control approaches like MPC, as gradients and solution sensitivities are ill-defined on the boundaries, and transitioning between mutually exclusive dynamics is nontrivial~\cite{bemporad1999,liberzon2003switching,Nurkanovic2022}.
To obtain a well-posed model at the boundaries, we use the Filippov differential inclusion (FDI)~\cite{filippov1960}
\begin{equation}
    \begin{split}
        \dot{x}\in F(x,u):=\Big\{\sum_{i=1}^m \theta_i f_i(x,u)\ \big|\ \sum_i\theta_i=1,\\
        \theta_i\ge0,\ \theta_i=0\ \text{if }x\notin\overline{R}_i\Big\}.\label{eq:filippov}
    \end{split}
\end{equation}
Here $\theta_i$ are functions used as convex multipliers for $F:\mathbb{R}^{n_x}\times\mathbb{R}^{n_u} \rightrightarrows \mathbb{R}^{n_x}$.
In the interior of $R_i$, the dynamics reduce to $F(x,u)=\{f_i(x,u)\}$. On boundaries, $F(x,u)$ is the convex combination of vector fields of the neighboring regions $R_i$ \cite{filippov1960,Nurkanovic2022}. This recasting of the dynamics as the FDI \eqref{eq:filippov} yields a well-posed continuous-time model. This model can be used to formulate a non-smooth Optimal Control Problem (OCP) and derive a real-time MPC scheme tailored to this structure.


\section{Non-Smooth MPC via NOSNOC}
\label{sec:nosnoc_mpc}

\subsection{MPC formulation}
Given the Filippov model $\dot x \in F(x,u)$, we pose the continuous-time OCP on a horizon $[0,T],\ T>0$~\cite{Nurkanovic2022}:
\begin{subequations}\label{eq:ocp}
\begin{align}
    \min_{x(\cdot),u(\cdot)}\quad &
    \int_{0}^{T} \ell(x(t),u(t))\,dt \;+\; \ell_\mathrm{T}(x(T)) \label{eq:ocp_ct_obj}\\
    \text{s.\,t.}\quad &
    x(0)=\hat{x}(t_s), \label{eq:ocp_init_dyn}\\
    & \dot x(t) \in F(x(t),u(t)), ~~ t \in [0,T], \label{eq:ocp_cont_dyn}\\
    & G(x(t),u(t))\le 0, ~~~~~ t \in [0,T], \label{eq:ocp_path_and_box_cons}\\
    & G_\mathrm{T}(x(T)) \le 0. \label{eq:ocp_term_cons}
\end{align}
\end{subequations}
Here, $\ell:\mathbb{R}^{n_x}\times\mathbb{R}^{n_u} \to \mathbb{R}$ is the stage cost, $\ell_\mathrm{T}:\mathbb{R}^{n_x} \to\mathbb{R}$ the terminal cost, and $G: \mathbb{R}^{n_x}\times\mathbb{R}^{n_u} \to \mathbb{R}^{n_G},G_\mathrm{T}: \mathbb{R}^{n_u} \to \mathbb{R}^{n_{G_\mathrm{T}}}$ collect path and terminal constraints. 
In the initial constraint \eqref{eq:ocp_init_dyn} $\hat{x}(t_s)$ is the real state value obtained at sampling instance $t_s$. To apply MPC, \eqref{eq:ocp} needs to be solved at sampling instants $t_s$ over a receding time horizon. 
Then, the first part of a control input $u(t), \ t \in [t_s,t_{s+1})$ is applied until $t_{s+1}$, and the process is repeated. 
In a discrete-time setting, the control input is usually piecewise constant. Thus, we compute $u(t) = u_s, \ t \in [t_s,t_{s+1})$, cf. Sec.~\ref{sec:discrete_time_mpc}.
To use \eqref{eq:ocp} for MPC, however, some reformulations are necessary.

\subsection{Dynamic Complementarity System}
To obtain a computationally efficient model of the Filippov inclusion, it is reformulated into a dynamic complementarity system (DCS) and discriminant functions $g~=~[g_i,\hdots,g_m],\, g_i:\mathbb{R}^{n_x}\to\mathbb{R}$ that encode the regions~\cite{stewart1990}:
\begin{equation}
R_i=\{\,x\mid g_i(x)<\min_{\substack{j \in \{1,\hdots,m\}\\j\neq i}}g_j(x)\,\}.
\end{equation}
At any $(x,u)$, the Filippov convex multipliers $\theta\in\mathbb{R}^m_{\ge 0}$ are the solution of the selector LP with $\theta = (\theta_i,\hdots,\theta_m)$~\cite{stewart1990,Nurkanovic2024}:
\begin{equation}
\min_{\theta}\; g(x)^\top \theta\quad\text{s.\,t.}\quad \mathbf{1}^\top\theta=1,\;\theta\ge 0.
\end{equation}
with $\mathbf{1} = [1,1,\hdots,1] \in \mathbb{R}^m$. A detailed derivation of this process is presented in~\cite{Nurkanovic2024}.
Introducing Lagrange multipliers $\lambda\in\mathbb{R}^m_{\ge 0}$ and $\mu\in\mathbb{R}$, the KKT conditions are
\begin{subequations}\label{eq:DCS_kkt}
\begin{align}
& g(x)-\lambda-\mu\textbf{1}=0,\\ 
& 1-\mathbf{1}^\top\theta=0,\\
& \theta\ge 0,\ \mu\ge 0, \theta \perp \mu. \label{eq:DCS_kkt_comp}
\end{align}
\end{subequations}
Together with 
\begin{equation}
\dot{x} = \sum_{i=1}^m \theta_i f_i(x,u),\label{eq:DCS_sys}
\end{equation}
this defines a DCS in $(x,\theta,\lambda,\mu)$ that is equivalent to the Filippov model in \eqref{eq:filippov}.
To facilitate the implementation of an MPC controller, we can now replace \eqref{eq:ocp_cont_dyn} in the OCP by the DCS~\eqref{eq:DCS_kkt}\,--\,\eqref{eq:DCS_sys}.
Afterwards, we discretize this OCP in time to obtain a finite-dimensional MPCC, which we can efficiently solve using NOSNOC.

\subsection{Time Discretization and Switch Detection}
To discretize the DCS~\eqref{eq:DCS_kkt}\,--\,\eqref{eq:DCS_sys}, we utilize the scheme presented in~\cite{Nurkanovic2024}. 
For ease of exposition, we consider a single control interval $[T_k,T_{k+1}] \subseteq [0,T]$ with a constant input $\bar{u}$ and initial value $x_0$.
We split it into $N_{\mathrm{FE}}$ finite elements $[t_n,t_{n+1}]$, with grid points $T_k=t_0<\dots<t_{N_{\mathrm{FE}}}=T_{k+1}$, with step sizes $h_n=t_{n+1}-t_n$.
On each element, we apply an $n_{\mathrm{RK}}$-stage Runge–Kutta (RK)~\cite{Nurkanovic2022}.
Let $x_n\approx x(t_n)$ be the left state, $V_n=(v_{n,1},\dots,v_{n,n_{\mathrm{RK}}})$ the RK-stage derivatives,
$\Theta_n=(\theta_{n,1},\dots,\theta_{n,n_{\mathrm{RK}}})$ the stage values of the selector,
and $(\Lambda_n,M_n)$ the corresponding Lagrange multipliers in~\eqref{eq:DCS_kkt}. 
We collect all RK-stage variables for the $n$-th finite element  in $Z_n=(x_n,V_n,\Theta_n,\Lambda_n,M_n)$.

For one element, we can summarize all RK-equations in $G_{\mathrm{rk}}$, which then stacks all RK equations for the DCS. 
Thus, over $[T_k,T_{k+1}]$, we stack $Z=(Z_0,\ldots,Z_{N_{\mathrm{FE}}-1})$ and ${H=(h_0,\ldots,h_{N_{\mathrm{FE}}-1})}$ to obtain
\begin{equation}
x_{k+1}=F_{\mathrm{std}}(Z),\qquad
0=G_{\mathrm{std}}(Z,H,x_k,\bar u),
\end{equation}
a discrete-time non-smooth map. Where $G_\mathrm{std}$ stacks all computations of the standard RK method over all integration intervals. 
This discretization is only accurate if the grid points and switching points intersect, which is rare in practice~\cite{acary2008}.

To address this, switch detection is introduced by making the step sizes $h_n$ optimization decision variables and augmenting $G_{\mathrm{std}}$ with cross-complementarity constraints $0=G_{\mathrm{cross}}(Z,H, x_k)$ and step equilibration $0=G_{\mathrm{eq}}(Z, H, x_k)$. This enforces a constant active set (i.e., switching mode) within each element by making changes only possible at grid points while keeping the elements equidistant from switching points.
For details on the functions $G_\mathrm{rk},\,G_\mathrm{std},\,G_\mathrm{cross}$ and $G_\mathrm{eq}$, cf.~\cite[Sec. III]{Nurkanovic2022}.

With the length constraint $\sum_{n=0}^{N_{\mathrm{FE}}-1} h_n=T_{k+1}-T_k$, the full finite element step with switch detection is~\cite{Nurkanovic2024}
\begin{equation}\label{eq:filippov_fesd}
\begin{aligned}
& x_{k+1}=F_{\mathrm{fesd}}(Z), \;
0=G_{\mathrm{fesd}}^k(Z,H,x_k,\bar u),\\
& G_{\mathrm{fesd}}^k(Z,H,x_k,\bar u):=
\begin{bmatrix}
G_{\mathrm{std}}(Z,H,x_k,\bar u)\\
G_{\mathrm{cross}}(\Theta,\Lambda)\\
G_{\mathrm{eq}}(H,\Theta,\Lambda)\\
\sum_{n=0}^{N_{\mathrm{FE}}-1} h_n - T_{k+1}+T_k
\end{bmatrix},
\end{aligned}
\end{equation}
which serves as an integrator with exact switch detection in a time-discretized OCP, described in the next section.

\subsection{Discrete-time MPC subproblem as an MPCC}
\label{sec:discrete_time_mpc}
To write the discretized version of the OCP \eqref{eq:ocp}, we consider $N_{\mathrm{CI}}$ control intervals of equal length and define  ${\mathbf{x}} = (x_0,\hdots,  x_{N_\mathrm{CI}})$ be the states at control grid nodes, $\mathbf{u} = (u_0,\hdots,u_{N_\mathrm{CI}-1})$ the control inputs, ${\mathcal{Z} = (Z_0,\hdots,Z_{N_\mathrm{CI}-1})}$ the collection of all internal variables, and $\mathcal{H} = (H_0,\hdots,H_{N_\mathrm{CI}-1})$ all step sizes stacked per stage as introduced above. 
Thus, the time-discrete version of the OCP~\eqref{eq:ocp} can now be written as
\begin{subequations}\label{eq:MPC_MPCC}
\begin{align}
\min_{{\mathbf{x}},\mathbf{u},\mathcal{Z},\mathcal{H}}\quad & \sum_{k=0}^{N_{\mathrm{CI}}-1} \hat\ell_k(x_k,Z_k, H_k,u_k) + \hat \ell_T(x_{N_{\mathrm{CI}}})\\
\text{s.\,t.}\quad &
{x}_0=x(t_s),\\
& G_T(x_{N_{\mathrm{CI}}}) \leq 0,\\ 
& \forall k \in (0,\hdots,N_\mathrm{CI}-1): \nonumber\\
& ~~~~  x_{k+1}=F_{\mathrm{fesd}}({Z}_k),\\
& ~~~~ G_{\mathrm{fesd}}^k(Z_k, H_k,{x}_k,u_k) = 0,\\
& ~~~~ G(x_k,u_k) \leq 0, \label{eq:MPC_MPCC_path}
\end{align}
\end{subequations}
where $\hat \ell_k$ and $\hat \ell_T$ define the time-discretized cost~\eqref{eq:ocp_ct_obj}.
The FDI~\eqref{eq:ocp_cont_dyn} is replaced by its time-discretization~\eqref{eq:filippov_fesd}, and the path constraints~\eqref{eq:ocp_path_and_box_cons} are evaluated at the control grid points yielding~\eqref{eq:MPC_MPCC_path}.
Because $G_{\mathrm{fesd}}$ embeds complementarity from the KKT conditions of the selector LP (cf.~\eqref{eq:DCS_kkt_comp}) and cross-complementarity constraints,
\eqref{eq:MPC_MPCC} is a mathematical program with complementarity constraints, which can be written in compact form
\begin{subequations}
\label{eq:mpcc_compact}
    \begin{align}
        \min_{w} f(w) \\
        \text{s.\,t.}\quad & h(w)\geq 0,\\
        & 0\leq w_1\ \perp w_2\geq 0, 
    \end{align}
\end{subequations}
for a suitable partition $w=(w_0,w_1,w_2) \in \mathbb{R}^{n_w}$ of all decision variables~\cite{Nurkanovic2022}.
We solve the MPCC via a homotopy, whereby NOSNOC supports, replacing the bilinear orthogonality by $w_1^\top w_2=\sigma_i$ (smoothing) or by $w_1^\top w_2\le \sigma_i$ (relaxation) and solving a sequence of smooth NLPs with a parameter $\sigma$ driven to zero. 
Under standard assumptions, letting $\sigma_i\rightarrow 0$ recovers an MPCC solution~\cite{Hoheisel2013}.


\section{Experimental Evaluation}
\label{sec:results}

\begin{figure}
    \centering
    \includegraphics[width=0.95\linewidth]{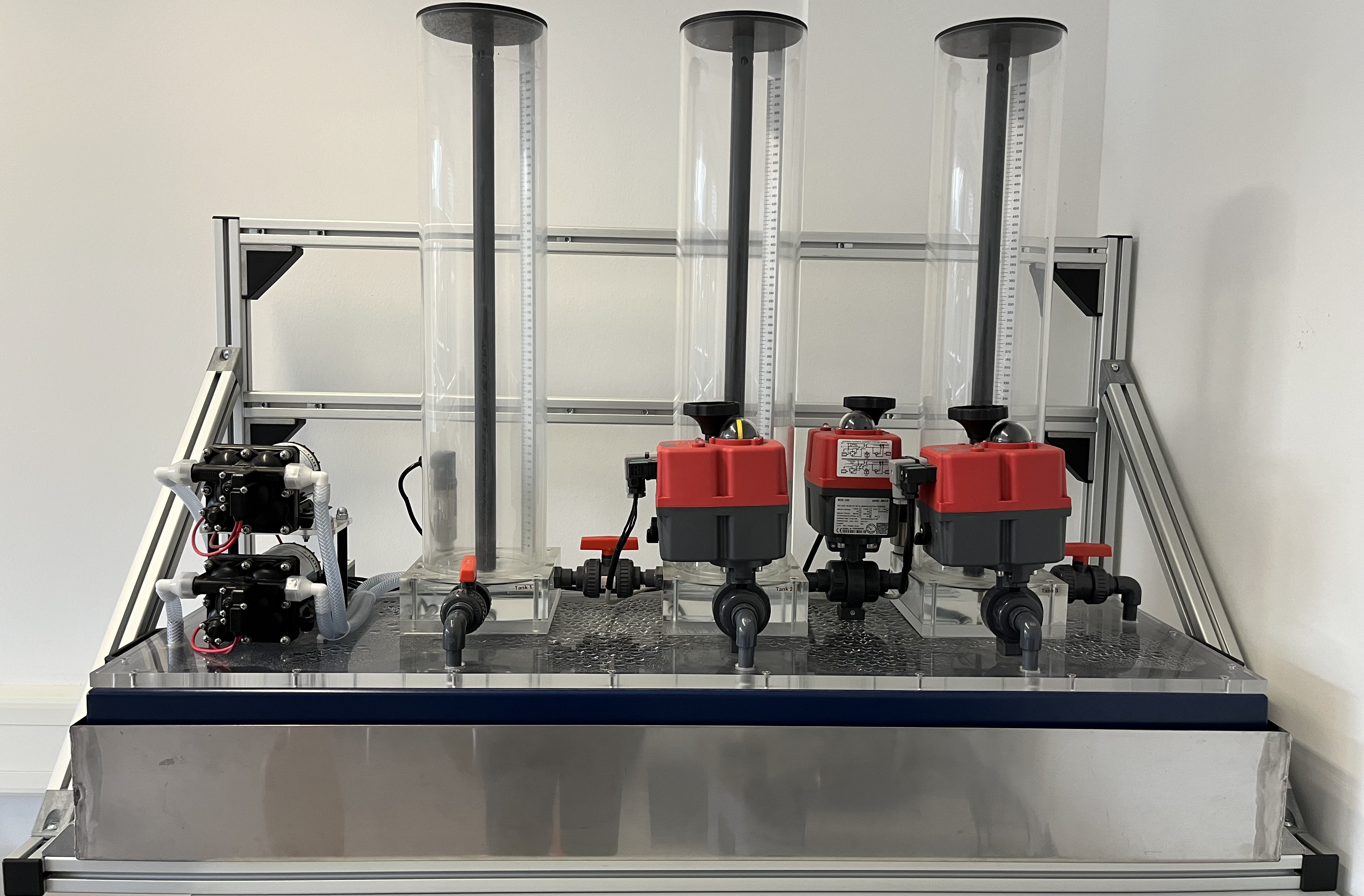}
    \vspace{-0.1cm}
    \caption{Image of the three-tank-system used in the experiments}
    \label{fig:three_tank_sys}
    \vspace{-0.55cm}
\end{figure}

To demonstrate our approach in real-time, we consider the three-tank system shown in Fig.~\ref{fig:three_tank_sys}. This setup consists of three similarly sized tanks and two pumps, feeding into the top of tanks~1 and 3. Each tank is equipped with a bottom-mounted level sensor and drains to the outside of the system. Furthermore, there are connections between tanks~1 and 2, as well as tanks~2 and 3. Each of these connections and drains is equipped with a continuous differential valve. The tank base area $A$ as well as all cross-sectional areas $q_i$, with $i\in\{1,2,\mathrm{d1},\mathrm{d2},\mathrm{d3}\}$ are denoted in Table \ref{tab:diameters} and can be changed between these values and 0 continuously via the valves. The parameters with subscripts $\mathrm{d}$ are drains, and the others are tank connections. The full piecewise-smooth model of the system can thus be written as in \eqref{eq:tank_pss}, 
\begin{figure*}
\vspace{0.2cm}
\hrulefill
\begin{subequations}
\label{eq:tank_pss}
\begin{align}
    \dot x_1 &= \frac{1}{A} \begin{cases} 
      u_1 - q_1\sqrt{2g(x_1-x_2)}-q_\mathrm{d1}\sqrt{2gx_1}, & \hspace{2.9cm} x_1 > x_2, \\
      u_1 + q_1\sqrt{2g(x_2-x_1)}-q_\mathrm{d1}\sqrt{2gx_1}, & \hspace{2.9cm} x_1 < x_2,
   \end{cases}\label{eq:tank1_pss}\\
   \dot x_2 &= \frac{1}{A} \begin{cases}
       q_1\sqrt{2g(x_1-x_2)} + q_2\sqrt{2g(x_3-x_2)} - q_\mathrm{d2}\sqrt{2gx_2}, & \hspace{5mm} x_1 > x_2 \wedge x_3 > x_2, \\
       -q_1\sqrt{2g(x_2-x_1)} + q_2\sqrt{2g(x_3-x_2)} - q_\mathrm{d2}\sqrt{2gx_2}, & \hspace{5mm} x_1 < x_2 \wedge x_3 > x_2, \\
       q_1\sqrt{2g(x_1-x_2)} - q_2\sqrt{2g(x_2-x_3)} - q_\mathrm{d2}\sqrt{2gx_2}, & \hspace{5mm} x_1 > x_2 \wedge x_3 < x_2, \\
       - q_1\sqrt{2g(x_2-x_1)} - q_2 \sqrt{2g(x_2-x_3)} - q_\mathrm{d2}\sqrt{2gx_2}, & \hspace{5mm} x_1 < x_2 \wedge x_3 < x_2,
       \end{cases}\label{eq:tank2_pss}\\
   \dot x_3 &= \frac{1}{A}\begin{cases}
       u_2 - q_2\sqrt{2g(x_3-x_2)} - q_\mathrm{d3}\sqrt{2gx_3}, & \hspace{2.9cm} x_3 > x_2, \\
       u_2 + q_2\sqrt{2g(x_2-x_3)} - q_\mathrm{d3}\sqrt{2gx_3}, & \hspace{2.9cm} x_3 < x_2,
       \end{cases}\label{eq:tank3_pss}
\end{align}
\end{subequations}
\hrulefill
\vspace{-0.2cm}
\end{figure*}
where $g=\qty{9.81}{\m \s^{-2}}$ is the gravitational acceleration, the heights are the states $x=[x_1,x_2,x_3]^\top$. In \eqref{eq:tank_pss}, the equal-hight sets ${x_1 = x_2}$ and ${x_2 = _3}$ are part of the switching set and thus pose a challenge regarding the square root $f_\mathrm{sqrt}(x)=\sqrt{x}$, which are defined only for $x \geq 0$ and are not Lipschitz at $x=0$.
Consequently, $\,f_\mathrm{sqrt}'(x)=\tfrac{1}{2\sqrt{x}}\, $ and $\, f_\mathrm{sqrt}''(x)=-\tfrac{1}{4}x^{-3/2}\,$ blow up as $x\rightarrow 0$,
which yields unbounded derivatives near the equal-level sets at the region boundaries.
A low complexity solution is to add regions symmetrically around the borders $\partial R_i$ and approximate the behavior linearly with e.g. $f_\mathrm{lin,sqrt} = q\sqrt{\frac{2g}{k}}(x_2 - x_1)$, where $k := \lVert x_1 - x_2 \rVert$ is the euclidean distance to the border. The larger the chosen $k$, the larger the model mismatch becomes. Adding these regions around the borders increases the number of regions from 4, as seen in \eqref{eq:tank_pss} to 9 if we only consider the connections and not include the drains.
We do this to avoid the proliferation of modes (here $m=3^2\cdot 2^3=72$), which would greatly inflate problem size. To remedy this, we can shift the feasible set away from the scenarios where the root-terms of the drains approach 0, by adjusting the lower bound to, e.g., $\qty{0.1}{\cm}$.
As a further safeguard, we introduce $\max$ functions under the remaining square-root terms to ensure positivity, i.e., clipping its argument to some minimum value $e$. This is done to prevent numerical issues in evaluating the dynamic equations.  These states need to be derived from the voltage signals produced by the level sensors, which is done by the following equations
\begin{align*}
x_1 &= \qty{29.942}{\cm} - \qty{3.344}{\cm \volt^{-1}} \cdot U_\mathrm{1,s},\\
x_2 &= \qty{29.449}{\cm} - \qty{3.326}{\cm \volt^{-1}} \cdot U_\mathrm{2,s},\\
x_3 &= \qty{28.948}{\cm} - \qty{3.330}{\cm \volt^{-1}} \cdot U_\mathrm{3,s},
\end{align*}
where $U_\mathrm{1,s},\,U_\mathrm{2,s}$, and $U_\mathrm{3,s}$ are the voltages output by the sensors. The control inputs $[u_1,u_2]$ are the flow rates driven by the pumps, which are themselves voltage-controlled. Therefore, the inputs are defined implicitly by the equations 
\begin{align*}
u_1 &= \qty{64.672}{\cm^3\s^{-1}} + \qty{6.186}{\cm^3\volt^{-1}\s^{-1}} \cdot U_\mathrm{1,p},\\
u_2 &= \qty{64.672}{\cm^3\s^{-1}} + \qty{6.186}{\cm^3\volt^{-1}\s^{-1}} \cdot U_\mathrm{2,p},
\end{align*}
where $U_\mathrm{1,p}$ and $U_\mathrm{2,p}$ are the input voltages of the pumps.

The control scheme is implemented via MATLAB 2023b using NOSNOC~\cite{Nurkanovic2022} for modeling the system, formulating and discretizing the OCP, and solving the resulting MPCC in an MPC loop.
NOSNOC uses CasADi~\cite{casadi} for generating the necessary functions used to solve the OCP. 
We use the Data Acquisition Toolbox in order to connect to the \textit{NI USB 6341} I/O card~\cite{iocard}, which is used for the interface to the physical system.
After installing the appropriate support package, the card was configured in MATLAB, defining its inputs, outputs, and sampling rate. 
The signal ranges of the card and the amplifier are limited to $[\qty{-10}{\volt},\qty{10}{\volt}]$. 
For the controller, we use the bounds shown in Table \ref{tab:bounds}.
We use a state-dependent quadratic stage cost function $\ell(x) = x^\top Qx$, where $Q$ is the identity matrix, and no terminal cost, with a prediction horizon $T=\qty{50}{s}$, with $N_{\mathrm{CI}} = 10$ control intervals and a second-order Runge-Kutta method. 
The number of finite elements $N_\mathrm{FE}$ was chosen as $2$. 
For the homotopy, we chose a superlinear update rule with 
$(\sigma_1,\sigma_2,\sigma_3,\sigma_4)=(1,10^{-1},10^{-3},10^{-9})$.
Further, we warm start the solver with a shifted form of the previous solution.
\begin{table}[t]
\centering
\caption{Values of the tank base area $A$ and all cross-sectional areas $q_i$}
\vspace{-0.15cm}
\begin{tabular}{|c|c|}
\hline
\textbf{Parameter} & \textbf{Cross-sectional area (cm\textsuperscript{2})} \\ \hline
$A$     & 153.941    \\ \hline
$q_{1}$ & 0.218 \\ \hline
$q_{2}$ & 0.228 \\ \hline
$q_{d1}$ & 0.373 \\ \hline
$q_{d2}$ & 0.427 \\ \hline
$q_{d3}$ & 0.331 \\ \hline
\end{tabular}
\vspace{-0.1cm}
\label{tab:diameters}
\end{table}
\begin{table}[t]
\centering
\caption{Values of the state and input bounds}
\vspace{-0.15cm}
\begin{tabular}{|c|c|c|}
\hline
\textbf{Bound} & \textbf{Lower} & \textbf{Upper} \\ \hline
$x_1,x_2,x_3$ & \qty{0.1}{\cm} & \qty{60}{\cm} \\ \hline
$u_1$ & \qty{0}{\cm^{3}\s^{-1}} & \qty{125}{\cm^{3}\s^{-1}} \\ \hline
$u_2$ & \qty{0}{\cm^{3}\s^{-1}} & \qty{137}{\cm^{3}\s^{-1}} \\ \hline
\end{tabular}
\label{tab:bounds}
\vspace{-0.35cm}
\end{table}
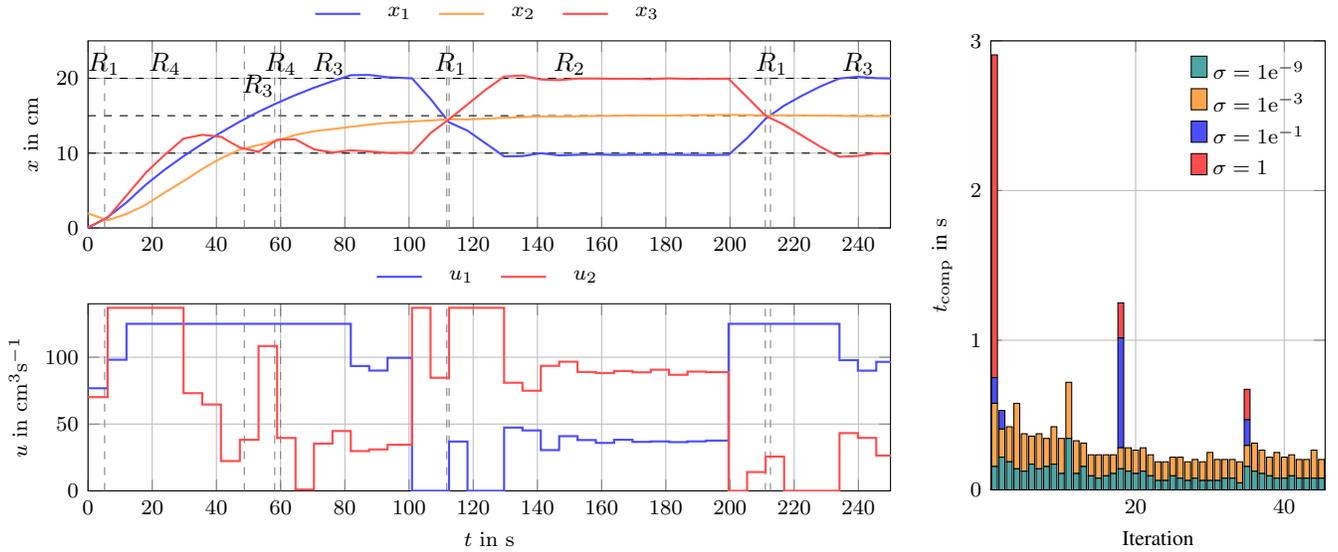
\begin{figure*}[t!]
    \centering
         \begin{tikzpicture}[scale=1]

\newlength\LeftHeight
\setlength{\LeftHeight}{0.14\textwidth}
\def\LeftWidth{0.6\textwidth}
\def\RightWidth{0.25\textwidth}
\def\HGap{0.075\textwidth}

\pgfplotsset{
  compat=1.18,
  every axis/.append style={
    scale only axis,     
    grid=both,
    ticklabel style={font=\footnotesize},
    xlabel style={font=\footnotesize},
    ylabel style={font=\footnotesize},
    legend style={font=\footnotesize, cells={align=left}},
    grid=both
  }
}

\begin{axis}[
  name=states,
  width=\LeftWidth,
  height=\LeftHeight,
  ylabel={$x~\mathrm{in~cm}$},
  xmin=0, xmax=250,
  ymin=0, ymax=25,
  legend columns=3,
  legend style={
    font=\footnotesize,
    column sep=8pt,
    draw=none,
    at={(0.5,1.05)},
    anchor=south,
  },
]
\addplot[black, dashed, forget plot] table[]{
0 20
290 20
};
\addplot[black, dashed, forget plot] table[]{
0 15
290 15
};
\addplot[black, dashed, forget plot] table[]{
0 10
290 10
};

\addplot[gray, dashed, forget plot] table[]{
5.1707 0
5.1707 30
};
\node at (axis cs:5,22) {$R_1$};

\addplot[gray, dashed, forget plot] table[]{
48.69 0
48.69 30
};
\node at (axis cs:24,22) {$R_4$};

\addplot[gray, dashed, forget plot]table[]{
58.1614 0
58.1614 30
};
\node at (axis cs:53,19) {$R_3$};

\addplot[gray, dashed, forget plot]table[]{
59.97 0
59.97 30
};
\node at (axis cs:60,22) {$R_4$};

\addplot[gray, dashed, forget plot] table[]{
111.76 0
111.76 30
};
\node at (axis cs:75,22) {$R_3$};

\addplot[gray, dashed, forget plot] table[]{
112.49 0
112.49 30
};
\node at (axis cs:113,22) {$R_1$};

\addplot[gray, dashed, forget plot] table[]{
211.02 0
211.02 30
};
\node at (axis cs:150,22) {$R_2$};

\addplot[gray, dashed, forget plot] table[]{
212.61 0
212.61 30
};
\node at (axis cs:213,22) {$R_1$};

\node at (axis cs:240,22) {$R_3$};

\addplot[thick,blue!70] table[] {
0 0.01
6.1068 1.4399
11.9869 3.4082
18.0313 5.7934
23.9193 7.7973
29.7588 9.5906
35.5958 11.2393
41.4093 12.7634
47.305 14.2119
53.1218 15.5287
58.914 16.6954
64.6494 17.7644
70.3887 18.7353
76.1475 19.6068
81.8658 20.4126
87.5915 20.4483
93.3107 20.1406
100.9331 19.9922
106.7138 17.2312
112.4632 14.1194
118.225 12.9984
123.9441 11.2786
129.625 9.5718
135.3403 9.6039
141.0627 9.991
146.7829 9.7016
152.4922 9.7782
158.2064 9.8234
163.8953 9.7557
169.6337 9.7843
175.3322 9.7804
181.0075 9.796
186.7557 9.7673
192.4739 9.7456
199.5434 9.776
205.3368 11.9968
211.1074 14.5998
216.8598 16.3462
222.6091 17.7143
228.3622 18.9477
234.1025 19.9737
239.8399 20.2009
245.5628 20.0156
251.3543 19.9737
257.0804 19.9863
262.8109 19.9632
268.5442 20.0213
274.2678 19.98
279.9831 19.9998
285.717 20.0479
291.4424 20.0281
};
\addlegendentry{$x_1$};

\addplot[thick,orange!70] table[] {
0 1.9414
6.1068 1.025
11.9869 1.8574
18.0313 3.1023
23.9193 4.7429
29.7588 6.2586
35.5958 7.8922
41.4093 9.2961
47.305 10.4646
53.1218 11.0774
58.914 11.6598
64.6494 12.441
70.3887 12.9286
76.1475 13.2102
81.8658 13.5126
87.5915 13.8123
93.3107 14.045
100.9331 14.2319
106.7138 14.3643
112.4632 14.5249
118.225 14.5
123.9441 14.608
129.625 14.7236
135.3403 14.8183
141.0627 14.8984
146.7829 14.9089
152.4922 14.9172
158.2064 14.9698
163.8953 14.9829
169.6337 15.0407
175.3322 15.017
181.0075 15.0133
186.7557 15.0657
192.4739 15.0858
199.5434 15.1099
205.3368 15.087
211.1074 15.0182
216.8598 15.041
222.6091 15.0383
228.3622 14.9994
234.1025 14.9858
239.8399 14.9312
245.5628 14.9449
251.3543 14.9445
257.0804 14.9469
262.8109 14.9381
268.5442 14.9535
274.2678 14.9159
279.9831 14.8931
285.717 14.8702
291.4424 14.8963
};
\addlegendentry{$x_2$};

\addplot[thick,red!70] table[] {
0 0.089951
6.1068 1.3634
11.9869 4.3424
18.0313 7.3892
23.9193 9.7811
29.7588 11.9443
35.5958 12.448
41.4093 12.1768
47.305 10.7463
53.1218 10.1787
58.914 11.7956
64.6494 11.8345
70.3887 10.5109
76.1475 10.0913
81.8658 10.3711
87.5915 10.2215
93.3107 10.0445
100.9331 10.0563
106.7138 12.6765
112.4632 14.5159
118.225 16.5022
123.9441 18.489
129.625 20.2329
135.3403 20.3707
141.0627 19.8393
146.7829 19.7537
152.4922 19.9603
158.2064 19.9667
163.8953 19.9222
169.6337 19.933
175.3322 19.8869
181.0075 19.9913
186.7557 19.9109
192.4739 19.9169
199.5434 19.9357
205.3368 17.4723
211.1074 14.9768
216.8598 13.802
222.6091 12.2995
228.3622 10.7956
234.1025 9.5124
239.8399 9.6247
245.5628 9.9841
251.3543 9.861
257.0804 9.7992
262.8109 9.8044
268.5442 9.814
274.2678 9.7718
279.9831 9.8715
285.717 9.8899
291.4424 9.8491
};
\addlegendentry{$x_3$};
\end{axis}

\begin{axis}[
  name=inputs,
  at={(states.below south west)}, anchor=above north west,
  width=\LeftWidth,
  height=\LeftHeight,
  ylabel={$u~\mathrm{in}~\mathrm{cm^3s^{-1}}$},
  xlabel={$t~\mathrm{in~s}$},
  xmin=0, xmax=250,
  ymin=0, ymax=140,
  legend columns=2,
  legend style={
    font=\footnotesize,
    column sep=8pt,
    draw=none,
    at={(0.5,1.05)},
    anchor=south,
  },
]

\addplot[gray, dashed, forget plot] table[]{
5.1707 0
5.1707 150
};
\addplot[gray, dashed, forget plot] table[]{
48.69 0
48.69 150
};
\addplot[gray, dashed, forget plot]table[]{
58.1614 0
58.1614 150
};
\addplot[gray, dashed, forget plot]table[]{
59.97 0
59.97 150
};
\addplot[gray, dashed, forget plot] table[]{
111.76 0
111.76 150
};
\addplot[gray, dashed, forget plot] table[]{
112.49 0
112.49 150
};
\addplot[gray, dashed, forget plot] table[]{
211.02 0
211.02 150
};
\addplot[gray, dashed, forget plot] table[]{
212.61 0
212.61 150
};

\addplot[thick, blue!70, const plot] table{
0 76.7966
6.1068 98.1277
11.9869 125
18.0313 125
23.9193 125
29.7588 125
35.5958 125
41.4093 125
47.305 125
53.1218 125
58.914 125
64.6494 125
70.3887 125
76.1475 125
81.8658 93.4509
87.5915 90.0169
93.3107 99.5636
100.9331 1.9193e-13
106.7138 1.0836e-13
112.4632 36.9302
118.225 8.6056e-13
123.9441 6.4037e-12
129.625 47.3633
135.3403 45.238
141.0627 30.5006
146.7829 40.9333
152.4922 38.0312
158.2064 35.8854
163.8953 38.2336
169.6337 36.6483
175.3322 37.0226
181.0075 36.47
186.7557 37.0387
192.4739 37.6416
199.5434 125
205.3368 125
211.1074 125
216.8598 125
222.6091 125
228.3622 125
234.1025 97.7899
239.8399 89.9519
245.5628 96.5448
251.3543 98.1073
257.0804 97.6347
262.8109 98.5587
268.5442 96.2887
274.2678 98.1533
279.9831 97.6189
285.717 96.0635
291.4424 96.5573
};
\addlegendentry{$u_1$}

\addplot[thick, red!70, const plot] table{

0 70.1618
6.1068 137
11.9869 137
18.0313 137
23.9193 137
29.7588 73.1025
35.5958 64.6386
41.4093 22.3242
47.305 38.2094
53.1218 108.3921
58.914 39.5898
64.6494 0.95093
70.3887 35.3904
76.1475 44.7832
81.8658 29.7497
87.5915 31.0037
93.3107 34.4988
100.9331 137
106.7138 84.6455
112.4632 137
118.225 137
123.9441 137
129.625 80.9248
135.3403 74.9385
141.0627 93.5615
146.7829 96.6478
152.4922 88.9775
158.2064 88.2058
163.8953 89.7168
169.6337 88.7249
175.3322 90.6586
181.0075 86.8647
186.7557 89.2797
192.4739 88.8558
199.5434 8.2128e-13
205.3368 14.0725
211.1074 25.6642
216.8598 5.7424e-13
222.6091 1.0421e-12
228.3622 2.5621e-11
234.1025 43.2268
239.8399 39.6528
245.5628 26.4502
251.3543 30.9499
257.0804 33.1757
262.8109 33.081
268.5442 32.5582
274.2678 34.4992
279.9831 31.0985
285.717 30.6611
291.4424 31.8756
};
\addlegendentry{$u_2$}

\end{axis}

\begin{axis}[
  name=comptime,
  at={($(states.north east)+(\HGap,0)$)}, anchor=north west,
  width=\RightWidth,
  height=2.4\LeftHeight,
  ybar stacked,
  bar width=2.2pt,
  ymin=0, ymax=3,
  xmin=0.5, xmax=45.5,
  ylabel={$t_\mathrm{comp}$ in $\mathrm{s}$},
  xlabel={Iteration},
  legend style={draw=none,fill=none},
]

\addplot[fill=teal!70, draw=black] table[]{
1 0.15625
2 0.21875
3 0.1875
4 0.14062
5 0.125
6 0.17188
7 0.14062
8 0.15625
9 0.17188
10 0.10938
11 0.34375
12 0.10938
13 0.15625
14 0.09375
15 0.078125
16 0.09375
17 0.10938
18 0.14062
19 0.125
20 0.10938
21 0.125
22 0.09375
23 0.0625
24 0.0625
25 0.09375
26 0.078125
27 0.0625
28 0.078125
29 0.0625
30 0.0625
31 0.0625
32 0.078125
33 0.078125
34 0.046875
35 0.15625
36 0.125
37 0.10938
38 0.09375
39 0.078125
40 0.078125
41 0.09375
42 0.078125
43 0.078125
44 0.078125
45 0.078125
46 0.0625
47 0.078125
48 0.078125
49 0.078125
50 0.078125
51 0.078125
};
\addlegendentry{$\sigma=1\mathrm{e}^{-9}$}

\addplot[fill=orange!70, draw=black] table[]{
1 0.42188
2 0.1875
3 0.23438
4 0.4375
5 0.25
6 0.1875
7 0.23438
8 0.1875
9 0.25
10 0.23438
11 0.375
12 0.21875
13 0.15625
14 0.14062
15 0.15625
16 0.14062
17 0.125
18 0.14062
19 0.15625
20 0.15625
21 0.15625
22 0.14062
23 0.125
24 0.125
25 0.125
26 0.14062
27 0.125
28 0.125
29 0.125
30 0.1875
31 0.14062
32 0.125
33 0.125
34 0.14062
35 0.14062
36 0.1875
37 0.15625
38 0.125
39 0.17188
40 0.14062
41 0.14062
42 0.125
43 0.125
44 0.1875
45 0.125
46 0.14062
47 0.14062
48 0.125
49 0.14062
50 0.14062
51 0.14062
};
\addlegendentry{$\sigma=1\mathrm{e}^{-3}$}

\addplot[fill=blue!70, draw=black] table[]{
1 0.17188
2 0.125
3 0
4 0
5 0
6 0
7 0
8 0
9 0
10 0
11 0
12 0
13 0
14 0
15 0
16 0
17 0
18 0.73438
19 0
20 0
21 0
22 0
23 0
24 0
25 0
26 0
27 0
28 0
29 0
30 0
31 0
32 0
33 0
34 0
35 0.17188
36 0
37 0
38 0
39 0
40 0
41 0
42 0
43 0
44 0
45 0
46 0
47 0
48 0
49 0
50 0
51 0
};
\addlegendentry{$\sigma=1\mathrm{e}^{-1}$}

\addplot[fill=red!70, draw=black] table[]{
1 2.1562
2 0
3 0
4 0
5 0
6 0
7 0
8 0
9 0
10 0
11 0
12 0
13 0
14 0
15 0
16 0
17 0
18 0.23438
19 0
20 0
21 0
22 0
23 0
24 0
25 0
26 0
27 0
28 0
29 0
30 0
31 0
32 0
33 0
34 0
35 0.20312
36 0
37 0
38 0
39 0
40 0
41 0
42 0
43 0
44 0
45 0
46 0
47 0
48 0
49 0
50 0
51 0
};
\addlegendentry{$\sigma=1~~~~~$}

\end{axis}

\end{tikzpicture}
         \vspace{-0.68cm}
        \caption{Scenario 1: Tank heights are shown on the top and inflows on the bottom. References show as dashed horizontal lines and region switches as dashed vertical lines. Detailed computation times are shown on the right (including  CPU times for each NLP solve in the MPCC homotopy).}
    \label{fig:scenario1pic}
    \vspace{-0.5cm}
\end{figure*}
\begin{figure*}
    \centering
    \begin{tikzpicture}[scale=1]

\setlength{\LeftHeight}{0.14\textwidth}
\def\LeftWidth{0.6\textwidth}
\def\RightWidth{0.25\textwidth}
\def\HGap{0.075\textwidth}

\pgfplotsset{
  compat=1.18,
  every axis/.append style={
    scale only axis,     
    grid=both,
    ticklabel style={font=\footnotesize},
    xlabel style={font=\footnotesize},
    ylabel style={font=\footnotesize},
    legend style={font=\footnotesize, cells={align=left}},
    grid=both
  }
}

\begin{axis}[
  name=states,
  width=\LeftWidth,
  height=\LeftHeight,
  ylabel={$x~\mathrm{in~cm}$},
  xmin=0, xmax=250,
  ymin=0, ymax=30,
  legend columns=3,
  legend style={
    font=\footnotesize,
    column sep=8pt,
    draw=none,
    at={(0.5,1.05)},
    anchor=south,
  },
]
\addplot[black, dashed, forget plot] table[]{
0 25
290 25
};
\addplot[black, dashed, forget plot] table[]{
0 20
290 20
};
\addplot[black, dashed, forget plot] table[]{
0 15
290 15
};

\addplot[gray, dashed, forget plot] table[]{
81.426 0
81.426 40
};
\node at (axis cs:40,5) {$R_4$};

\addplot[gray, dashed, forget plot] table[]{
117.4449 0
117.4449 40
};
\node at (axis cs:100,5) {$R_3$};

\addplot[gray, dashed, forget plot]table[]{
182.249 0
182.249 40
};
\node at (axis cs:150,5) {$R_2$};

\addplot[gray, dashed, forget plot]table[]{
186.046 0
186.046 40
};
\node at (axis cs:184,5) {$R_1$};
\node at (axis cs:220,5) {$R_3$};

\addplot[thick,blue!70] table[] {
0 0.01
6.1149 1.9971
12.0697 4.9347
17.9971 7.1362
23.8826 9.0149
30.0038 10.7389
36 12.432
44.2605 13.9897
50.1332 15.9728
55.9445 17.3356
61.6659 18.6068
67.4292 19.7983
73.1549 20.8941
78.8824 21.9045
84.9927 22.8286
91.2554 23.8214
92.0992 24.0055
97.8444 24.8098
104.8339 25.4013
112.6011 22.763
119.027 18.3762
124.7336 15.7354
130.444 14.3303
136.1063 14.9284
141.7852 15.0042
147.51 14.8365
153.1958 14.8351
158.9044 14.8232
164.5966 14.8353
171.6334 14.7984
178.161 16.7221
181.1208 17.3897
182.2132 18.6195
187.9283 20.4728
193.6626 21.7492
199.3989 22.8896
205.1143 23.8516
210.8147 24.7183
216.5414 25.4005
222.2624 25.3526
227.9707 25.1298
233.6739 25.1705
239.3505 25.209
245.0693 25.2436
250.7703 25.2798
};
\addlegendentry{$x_1$};

\addplot[thick,orange!70] table[] {
0 0.01
6.1149 0.23642
12.0697 1.3838
17.9971 2.8659
23.8826 4.4888
30.0038 6.1115
36 7.8434
44.2605 9.5645
50.1332 11.8855
55.9445 13.3767
61.6659 14.562
67.4292 15.5181
73.1549 16.2988
78.8824 17.1593
84.9927 17.9082
91.2554 18.4598
92.0992 18.5749
97.8444 19.0584
104.8339 19.2019
112.6011 19.2535
119.027 19.4391
124.7336 19.4506
130.444 19.4693
136.1063 19.6096
141.7852 19.7789
147.51 19.8808
153.1958 19.9673
158.9044 20.057
164.5966 20.1113
171.6334 20.1747
178.161 20.1542
181.1208 20.1252
182.2132 20.0538
187.9283 19.7148
193.6626 19.5888
199.3989 19.4167
205.1143 19.2997
210.8147 19.2936
216.5414 19.3644
222.2624 19.4144
227.9707 19.4476
233.6739 19.4842
239.3505 19.4998
245.0693 19.5378
250.7703 19.5617
};
\addlegendentry{$x_2$};

\addplot[thick,red!70] table[] {
0 0.01
6.1149 2.0686
12.0697 5.4163
17.9971 8.1452
23.8826 10.4854
30.0038 12.6249
36 14.7106
44.2605 16.5918
50.1332 18.1716
55.9445 16.7376
61.6659 16.1742
67.4292 15.7485
73.1549 16.5188
78.8824 17.432
84.9927 17.5007
91.2554 18.1961
92.0992 18.3303
97.8444 16.75
104.8339 15.0148
112.6011 16.807
119.027 20.3021
124.7336 22.5123
130.444 24.2194
136.1063 25.4425
141.7852 25.312
147.51 24.9426
153.1958 24.9158
158.9044 25.1291
164.5966 25.1342
171.6334 24.9862
178.161 22.2906
181.1208 21.4972
182.2132 19.9774
187.9283 18.2646
193.6626 16.5388
199.3989 14.8191
205.1143 14.8777
210.8147 15.2204
216.5414 15.0974
222.2624 15.0126
227.9707 15.015
233.6739 15.0179
239.3505 15.0859
245.0693 15.0199
250.7703 15.038
};
\addlegendentry{$x_3$};
\end{axis}

\begin{axis}[
  name=inputs,
  at={(states.below south west)}, anchor=above north west,
  width=\LeftWidth,
  height=\LeftHeight,
  ylabel={$u~\mathrm{in}~\mathrm{cm}^3\mathrm{s}^{-1}$},
  xlabel={$t~\mathrm{in~s}$},
  xmin=0, xmax=250,
  ymin=0, ymax=140,
  legend columns=2,
  legend style={
    font=\footnotesize,
    column sep=8pt,
    draw=none,
    at={(0.5,1.05)},
    anchor=south,
  },
]

\addplot[gray, dashed, forget plot] table[]{
81.426 0
81.426 140
};

\addplot[gray, dashed, forget plot] table[]{
117.4449 0
117.4449 140
};

\addplot[gray, dashed, forget plot]table[]{
182.249 0
182.249 140
};

\addplot[gray, dashed, forget plot]table[]{
186.046 0
186.046 140
};

\addplot[thick, blue!70, const plot] table{
0 125
6.1149 125
12.0697 125
17.9971 125
23.8826 125
30.0038 125
36 125
44.2605 125
50.1332 125
55.9445 125
61.6659 125
67.4292 125
73.1549 125
78.8824 125
84.9927 125
91.2554 0.45639
92.0992 125
97.8444 119.6264
104.8339 1.5635e-13
112.6011 2.0093
119.027 1.0739e-12
124.7336 21.0799
130.444 70.7418
136.1063 47.2109
141.7852 42.7927
147.51 48.0015
153.1958 47.2393
158.9044 46.8005
164.5966 45.8562
171.6334 125
178.161 125
181.1208 125
182.2132 116.664
187.9283 125
193.6626 125
199.3989 125
205.1143 125
210.8147 121.7177
216.5414 96.132
222.2624 97.4964
227.9707 105.3954
233.6739 103.5933
239.3505 102.0359
245.0693 100.4578
250.7703 98.9254
};
\addlegendentry{$u_1$}

\addplot[thick, red!70, const plot] table{

0 116.7808
6.1149 137
12.0697 137
17.9971 137
23.8826 137
30.0038 137
36 137
44.2605 104.7529
50.1332 29.2477
55.9445 66.0217
61.6659 54.0622
67.4292 94.3524
73.1549 89.6752
78.8824 59.2539
84.9927 77.3599
91.2554 48.126
92.0992 1.2552e-12
97.8444 4.7947e-12
104.8339 137
112.6011 124.9819
119.027 137
124.7336 137
130.444 128.3003
136.1063 81.964
141.7852 85.1239
147.51 97.7218
153.1958 97.8425
158.9044 89.0969
164.5966 88.3454
171.6334 2.5192e-13
178.161 1.0199e-13
181.1208 57.4402
182.2132 25.1271
187.9283 1.6643e-12
193.6626 8.1852e-12
199.3989 50.9513
205.1143 48.6643
210.8147 35.4826
216.5414 38.9951
222.2624 41.527
227.9707 41.1066
233.6739 40.5765
239.3505 37.9037
245.0693 39.8855
250.7703 38.9471
};
\addlegendentry{$u_2$}

\end{axis}

\begin{axis}[
  name=comptime,
  at={($(states.north east)+(\HGap,0)$)}, anchor=north west,
  width=\RightWidth,
  height=2.4\LeftHeight,
  ybar stacked,
  legend columns = 2,
  bar width=2.4pt,
  ymin=0, ymax=3,
  xmin=0.5, xmax=45.5,
  ylabel={$t_\mathrm{comp}$ in $\mathrm{s}$},
  xlabel={Iteration},
  legend style={draw=none,fill=none},
]

\addplot[fill=teal!70, draw=black] table[]{
1 0.1875
2 0.17188
3 0.1875
4 0.23438
5 0.14062
6 0.15625
7 0.20312
8 0.15625
9 0.15625
10 0.14062
11 0.14062
12 0.14062
13 0.14062
14 0.125
15 0.125
16 0.23438
17 0.10938
18 0.125
19 0.125
20 0.125
21 0.10938
22 0.078125
23 0.09375
24 0.078125
25 0.078125
26 0.10938
27 0.078125
28 0.21875
29 0.10938
30 0.10938
31 0.26562
32 0.21875
33 0.125
34 0.10938
35 0.09375
36 0.10938
37 0.10938
38 0.09375
39 0.10938
40 0.078125
41 0.078125
42 0.078125
43 0.078125
44 0.078125
45 0.078125
};
\addlegendentry{$\sigma=1\mathrm{e}^{-9}$}

\addplot[fill=orange!70, draw=black] table[]{
1 0.09375
2 0.125
3 0.28125
4 0.25
5 0.26562
6 0.5
7 0.32812
8 2.6094
9 0.125
10 0.20312
11 0.14062
12 0.17188
13 0.125
14 0.14062
15 0.53125
16 0.5625
17 0.125
18 0.14062
19 0.10938
20 0.14062
21 0.10938
22 0.15625
23 0.125
24 0.10938
25 0.125
26 0.14062
27 0.10938
28 0.10938
29 0.125
30 0.20312
31 0.79688
32 0.23438
33 0.125
34 0.125
35 0.15625
36 0.14062
37 0.14062
38 0.14062
39 0.125
40 0.125
41 0.125
42 0.1875
43 0.125
44 0.14062
45 0.125
};
\addlegendentry{$\sigma=1\mathrm{e}^{-3}$}

\addplot[fill=blue!70, draw=black] table[]{
1 0.20312
2 0.17188
3 0
4 0
5 0
6 0
7 0
8 0
9 0.125
10 0
11 0
12 0
13 0
14 0
15 0
16 0
17 0.21875
18 0
19 0.20312
20 2.0156
21 0.6875
22 0
23 0
24 0
25 0
26 0
27 0
28 0
29 0
30 0.15625
31 0
32 0.1875
33 0.45312
34 0
35 0
36 0
37 0
38 0
39 0
40 0
41 0
42 0
43 0
44 0
45 0
};
\addlegendentry{$\sigma=1\mathrm{e}^{-1}$}

\addplot[fill=red!70, draw=black] table[]{
1 0.51562
2 0
3 0
4 0
5 0
6 0
7 0
8 0
9 0
10 0
11 0
12 0
13 0
14 0
15 0
16 0
17 0
18 0
19 0.20312
20 0
21 0
22 0
23 0
24 0
25 0
26 0
27 0
28 0
29 0
30 0.21875
31 0
32 1.9531
33 0
34 0
35 0
36 0
37 0
38 0
39 0
40 0
41 0
42 0
43 0
44 0
45 0
};
\addlegendentry{$\sigma=1~~~~~$}

\end{axis}

\end{tikzpicture}
    \vspace{-0.7cm}
        \caption{Scenario 2: Tank heights are shown on the top and inflows on the bottom. 
        References show as dashed horizontal lines and region switches as dashed vertical lines. 
        Detailed computation times are shown on the right (including  CPU times for each NLP solve in the MPCC homotopy).}
    \label{fig:scenario2pic}
    \vspace{-0.5cm}
\end{figure*}
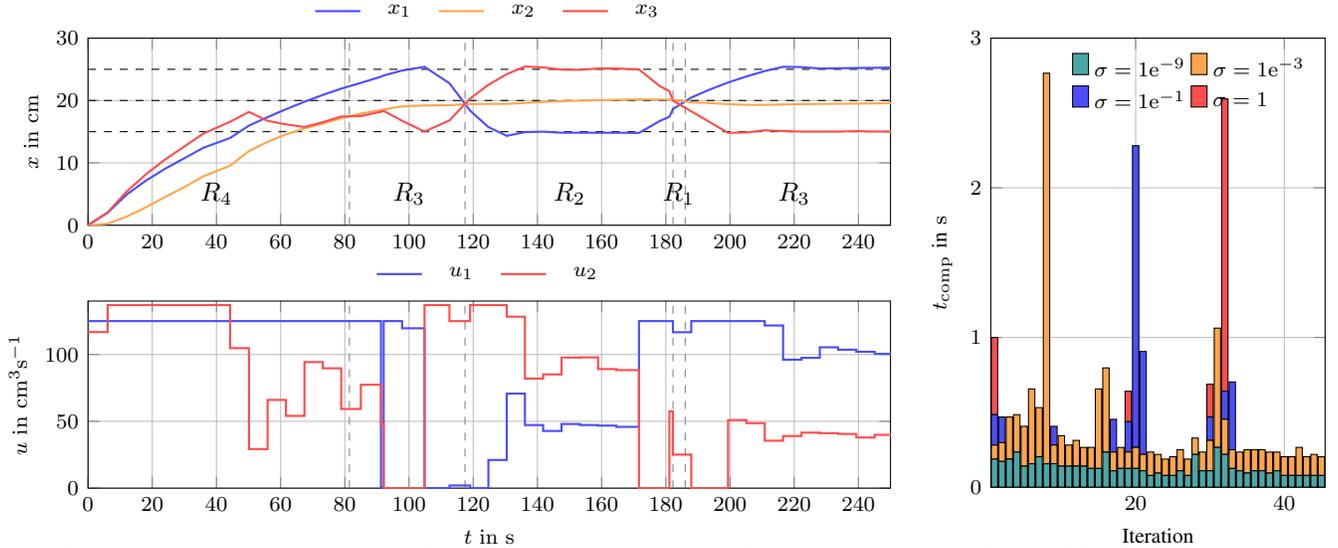

\subsection{Scenario Definition}

In all scenarios, our goal is to drive the system to a defined operating point. Note that not all operating points are reachable, as our system is underactuated, having no direct control for the level in tank~2. For the regions, we consider the regions of the real unaugmented model \eqref{eq:tank_pss} given as
\begin{equation*}
    \begin{aligned}
        R_1 &:= \{ x \in \mathbb{R}^3 \mid x_1 < x_2,\ x_2 > x_3 \},\\
        R_2 &:= \{ x \in \mathbb{R}^3 \mid x_1 < x_2,\ x_2 < x_3 \},\\
        R_3 &:= \{ x \in \mathbb{R}^3 \mid x_1 > x_2,\ x_2 > x_3 \},\\ 
        R_4 &:= \{ x \in \mathbb{R}^3 \mid x_1 > x_2,\ x_2 < x_3 \}. 
    \end{aligned}
\end{equation*}
The valve in the drain of tank~2 is closed in all scenarios, meaning $q_\mathrm{d2}=0$, and we chose $k=1$ and $e=1\mathrm{e}^{-6}$.
\paragraph{Scenario 1} Tank~1 should be driven to $\qty{20}{\cm}$, tank~2 to $\qty{15}{\cm}$, and tank 3 $\qty{20}{\cm}$. After $\qty{100}{\s}$, we switch the operating point values of tanks 1 and 3, and after $\qty{200}{\s}$, we switch them back.
\paragraph{Scenario 2} We investigate how well the MPC with the augmented model works for a model mismatch. For this, we use a similar setup as for Scenario 1 but reduce the diameter of the connection valves to $\qty{70}{\%}$.
The goal is to reach a operating point for tank 1 of $\qty{25}{\cm}$, tank 2 of $\qty{20}{\cm}$, and tank 3 of $\qty{15}{\cm}$, switching the operating points of tank 1 and 3 after $\qty{95}{\s}$ and $\qty{165}{\s}$.

\subsection{Scenario 1: Reference cascade}
\vspace{-0.05cm}
Figure~\ref{fig:scenario1pic} shows the closed-loop trajectories, references, and the computation times for each control interval. The system undergoes multiple mode transitions in the first control interval since a direct transition $R_1\!\to\! R_4$ is impossible in both the unaugmented model \eqref{eq:tank_pss} and the augmented model.
The tank levels rise to their setpoints, with only tank~1 showing an initial overshoot, which takes several control intervals to settle, but does not recur at the setpoint switches. At $t=\qty{100}{\s}$ the controller executes the mirrored setpoint (swapping tank~1 and~3), and at $t=\qty{200}{\s}$ it returns to the original target. Deviations occur mainly around the switching instant. Tank~2 remains at the setpoint despite the switching of the reference.
Computation times peak near switching events and are otherwise lower during near-steady operation.

\subsection{Scenario 2: Model-mismatch}
In Figure~\ref{fig:scenario2pic}, we show the state and input trajectories during the experimental run with model-mismatch. The tank levels rise from the initially empty tanks to the setpoints. Tank~3 shows a larger overshoot than in Scenario~1. Tank~2 does not settle to the set point and shows a small oscillating behavior around the reference.

As in the first scenario, the reference switches are performed by MPC with little over- and undershoot despite forcing region switches.
The computation time is also comparable and similarly peaks during setpoint switches. These peaks are more pronounced and span several control intervals. The homotopy iterations also remain at higher $\sigma$ values for several control intervals during these peaks.
However, in both cases the computation times stay below the sampling time of $\qty{5}{\s}$, rendering the MPC real-time feasible.

\subsection{Discussion}
The experiments indicate that non-smooth MPC achieves real-time capable control of a switching three-tank process while preserving mode consistency. Across scenarios, reference changes are tracked without unwanted behavior at region borders. Hence, the discrete dynamics remain faithful to the Filippov model and deliver reliable sensitivities.

Despite the multiple mode changes in the first control interval in Scenario~1, trajectories converge to the operating point. The brief overshoot of tank~1 vanishes at later setpoint swaps, consistent with improved warm starts and a closer proximity to the new targets, which reduces aggressive transients likely due to better active-set guesses. The overshoot might also be the result of the interconnected dynamics and might be needed to drive tank~2 to its reference point.

Because tank~2 is not directly actuated, the feasibility of operating points depends on the inter-tank flow balance. In Scenario~1, tank~2 settles at its target. Because of the connection mismatch in Scenario~2, tank~2 exhibits a small steady error and light oscillations, while tank~3 shows a larger overshoot. This could be addressed via an offset-free MPC scheme using a disturbance model~\cite{Morari2012,Pannocchia15}. Nevertheless, constraint satisfaction and correct mode sequencing are preserved, and the reference swaps are performed successfully.

To regularize non-Lipschitz square-root terms near zero, we combined a narrow linear region around the border with clipping via $\max(e,\cdot)$ and a small shift of the state lower bounds. This augmentation improves numerical robustness. The observed tracking quality suggests that, for the exercised operating points, this modified model is sufficient.

Computation times peak at setpoint changes and near major mode updates, then settle at lower values during near-steady operation. The peak can be explained by the new reference forcing a cold start, and that several homotopy steps are needed until the active set stabilizes, during which the $\sigma$ remains comparatively large. With the chosen schedule for $\sigma$ and warm starts of states, inputs, and multipliers, the controller remains real-time capable in both scenarios.


\section{Conclusion and Outlook}
\label{sec:conclusion} 
We presented a non-smooth MPC approach based on the NOSNOC framework, which combines Filippov/DCS modeling, FESD discretization, and homotopy-based MPCC solving. We validate its performance for switching systems on a real three-tank system.

The methodology maintains the switching behavior, aligns discretization with mode transitions, and avoids mixed-integer formulations. 
In the laboratory setup, the controller preserves mode sequencing, handles sliding and boundary behavior consistently, and achieves accurate reference tracking while satisfying constraints. 
Computation remains within real-time limits, and the approach shows robustness to structural model mismatch, indicating practical deployability of non-smooth MPC based on continuous optimization.

A limitation stems from the computational effort associated with solving MPCCs, which restricts achievable sampling rates and system complexity. 
Future directions should focus on accelerating the optimization pipeline through tailored MPCC solvers, improved homotopy strategies, and structure-exploiting numerical techniques. 
Additional extensions include applying the methodology to more complex non-smooth systems, such as mechanical contact or process systems with richer logic, and integrating offset-free, robust, or stochastic MPC formulations into the non-smooth setting.



\bibliographystyle{IEEEtran}
\bibliography{literature}

\end{document}